\pdfoutput=1
\documentclass[aps,american,pra,citeautoscript,floatfix,pra,pdftex,superscriptaddress,%
longbibliography,twocolumn,%
]{revtex4-1}
\usepackage{amsfonts,amsmath,amssymb}
\usepackage[T1]{fontenc}
\usepackage[utf8]{inputenc}
\usepackage{graphicx}
\usepackage{microtype}
\usepackage[obeyFinal,textsize=footnotesize]{todonotes}
\usepackage{stackrel}

\usepackage{natbib} 
\usepackage[normalem]{ulem}
\usepackage{graphicx}
\usepackage{amssymb}
\usepackage{amsmath}
\usepackage{bm}
\usepackage{bbm}
\usepackage{subfigure}
\usepackage{setspace}
\usepackage{hyperref}
\usepackage{color}
\hypersetup{%
  pdfpagemode=None, 
  pdfstartpage=1,
  pdfstartview=FitH,
  pdfmenubar=true,
  pdftoolbar=true,
  colorlinks = true,
  linkcolor=blue,
  citecolor=blue,
  bookmarksopen=false
}

\renewcommand{\equationautorefname}{Eq.}
\def\equationautorefname~#1\null{Eq. (#1)\null}

\usepackage{graphicx}  
\usepackage{dcolumn}   
\usepackage{bm}        
\usepackage{amssymb}   

\usepackage{cancel}
\usepackage{soul}

\usepackage{lmodern}

\newcommand{\expect}[1]{\langle{#1}\rangle}

\newcommand{\ie}{i.\,e.}%

\newcommand{\au}{a.u.}%

\graphicspath{{./Figures/}}

\begin{document}

\title{A linear polar molecule in a two-color cw laser field: a symmetry analysis}
\author{David Mellado-Alcedo}
\affiliation{Instituto Carlos I de F\'{\i}sica Te\'orica y Computacional,
and Departamento de F\'{\i}sica At\'omica, Molecular y Nuclear,
  Universidad de Granada, 18071 Granada, Spain} 
\author{Niurka R. Quintero}
\affiliation{Departamento de F\'{\i}sica Aplicada I, Universidad de Sevilla, 41011 Sevilla, Spain}
\author{Rosario Gonz\'alez-F\'erez}
\affiliation{Instituto Carlos I de F\'{\i}sica Te\'orica y Computacional,
and Departamento de F\'{\i}sica At\'omica, Molecular y Nuclear,
  Universidad de Granada, 18071 Granada, Spain}

\date{\today}
\begin{abstract}
A theoretical study of the rotational dynamics of a linear polar molecule in a two-color 
non-resonant cw laser field is presented. By systematically considering the interactions of this field with the electric dipole 
moment,  polarizability and hyperpolarizability of the molecule,  the effect of the symmetries of the Hamiltonian  
on the orientation and alignment  is explored in a regime where the  time-average approximation does not hold.  It
is shown that the alignment and orientation satisfy certain symmetries as a function of the phases and field strengths.
On average a one-color cw laser field doest not orient the molecule, being
necessary a two-color one having odd and even products of the laser  frequency to break 
the head-versus-tail order confinement.

\end{abstract}
\maketitle

\section{Introduction}
\label{sec:introduction}

Biharmonic signals are widely used in many areas of physics in order to break the time-shift symmetry 
of the external forces and of the electromagnetic 
fields~\cite{flach:2000,reimann:2002a,schiavoni:2003,ooi:2007,hanggi:2009,zolotaryuk:2011}. 
This symmetry  breaking  induces a plethora of unexpected phenomena, as for example, the 
dissipation-induced net motion, the current reversals by increasing  the amplitudes, and resonances as 
a function of the frequency and of the damping coefficient~\cite{cubero:2010,quintero:2011}. These 
phenomena have been observed in seemingly unrelated systems, such as 
semiconductors~\cite{seeger:1978}, Josephson junctions~\cite{ustinov:2004}, optical 
lattices~\cite{gommers:2005a}, ferrofluids~\cite{engel:2003}, Brownian particles~\cite{reimann:2002a}, 
Bose-Einstein condensates~\cite{salger:2009}, 
or solitons in  non-linear systems~\cite{salerno:2002,morales-molina:2003,quintero:2014}. 
Recent studies show that, regardless of the system, the symmetries of the biharmonic force determine
 the dependence of the measurements of the amplitudes and phases of this biharmonic 
force~\cite{quintero:2010,Cuesta,Casado}.

The spherical symmetry of a thermal sample of molecules is broken by inducing orientation and 
alignment~\cite{friedrich:nat,friedrich:zpd,friedrich:jpc95,stapelfeldt:2013} with experimental techniques such as brute force 
orientation~\cite{loesch:jcp93,bulthuis:jpca101,li:j_phys_chem_a}, 
combined electrostatic and  non-resonant laser  
fields~\cite{friedrich:jcp111,friedrich:jpca103-a,sakai:prl_90,Minemoto03,kupper:prl102,nielsen:prl2012,omiste:pra2012}, 
THz pulses~\cite{16,17,18,lapert:2012,shu:2013,egodapitiya:2014}, or the phase-locked 
two-color laser field~\cite{14,15,kraus:jpb,kraus:prl,Baykusheva:2016}. An aligned molecule is characterized by the confinement of the molecular 
fixed axes along the laboratory fixed frame, and keeping the head-versus-tail symmetry. 
For an oriented molecule, this symmetry is broken and the dipole moment is pointing towards one hemisphere rather than the opposite. 
In the spirit of biharmonic signals, continuous-wave non-resonant laser fields 
could be employed to create directional states of polar molecules, 
rather than the  laser pulses
used in experiments, whose time-envelope  is often given by a gaussian function.

Here, a linear polar molecule in a two-color continuous-wave non-resonant laser field is considered. 
Within the Born-Oppenheimer and the rigid-rotor approximations, the field-dressed rotational dynamics is 
analyzed.
The laser frequency is chosen so that the time-average approximation~\cite{Pershan} is not correct, but still 
assuming that no electronic, vibrational or rotational transitions are driven by this field.
By systematically including in the Hamiltonian the interactions of the field with the electric dipole moment, 
polarizability and hyperpolarizability, the effect of the symmetries of the system on  the 
orientation and alignment is analyzed.
Due to these symmetries, it is shown that it is not possible to orient on average 
the polar molecule with a one-color cw laser field being necessary to employ a two-color one, whose 
summed frequencies should be an odd multiple of the main laser frequency. 
In addition, the alignment and orientation satisfy certain  symmetries with respect to the 
relative phase of the two  electric  field components, and can be expressed as series expansions for
a fixed propagation time.  

The paper is organized as follow. In~\autoref{sec:hamiltonian}, the Hamiltonian of the system 
and its symmetries are described, and how they affect to the expectation values that characterize the field-dressed
rotational dynamics.
The results are analyzed and discussed in~\autoref{sec:results}. In~\autoref{subsec:time-average} 
the validity of the time-average approximation is investigated by including  only the interaction between the 
two-color electric field and the molecular dipole moment, and if higher order terms in 
the interactions are taken into account. The results for the 
orientation and alignment are analyzed in~\autoref{subsec:dip} and~\autoref{subsec:alig}, respectively,
by systematically including in the description the interactions of the electric field with the 
electric dipole moment, polarizability, and hyperpolarizability of the molecule.
The conclusions are given in~\autoref{sec:conclusions}.

\section{The System and the Hamiltonian}
\label{sec:hamiltonian}
A  linear polar molecule exposed to a phase-controlled continuous-wave (cw)
two-color laser field linearly polarized along the laboratory fixed frame (LFF)  $Z$-axis is considered. 
The corresponding electric field $\mathbf{E}(t)=E(t)\mathbf{Z}$ is given by the biharmonic function
\begin{equation}
E(t)=\sum_{i=1,2}\epsilon_i\cos\left[q_i\omega (t+t_0) + \delta_i\right], 
\label{eq:electric_field}
\end{equation}
with  $q_i \omega$, $\epsilon_i$ and $\delta_i$ being  the laser frequency, 
electric  field strength and  phase of the $i$-th harmonic, respectively, with 
$q_i$ being positive integers and $i=1,2$.
$t_0$ represents the time shift of the laser with respect 
to the initial time, \ie, when the molecule starts
to interact  with the two-color laser field, which was previously turned on, and, therefore $t_0$ 
might not be zero.

The molecule is described by the Born-Oppenheimer approximation, and the 
rotational motion is investigated using the rigid rotor approach. Within this framework, the Hamiltonian is given 
by~\cite{Hirschfelder,stone}
\begin{equation}
H=H_0+H_\mu+H_\alpha+H_\beta, 
\label{eq:molecular_Hamiltonian}
\end{equation}
where the first term stands for the  field-free Hamiltonian
\begin{equation}
H_0=\text{B}\mathbf{J}^2, \label{eq:field_free_Hamiltonian}
\end{equation}
with $B$ being the rotational constant of the molecule, and $\mathbf{J}$ the rotational angular momentum 
operator. The second, third and fourth  terms represent the interaction of the electric field with the 
electric dipole moment, polarizability, and hyperpolarizability of the molecule,  respectively, 
\begin{eqnarray}
H_\mu&=&-\mu\cos\theta\,E(t), \label{eq:dip_Hamiltonian}  \\
H_\alpha&=&-\frac1{2}\left(\Delta\alpha\cos^2\theta+\alpha_\bot\right)E^2(t),
\label{eq:pola_Hamiltonian} \\
H_\beta&=&-\frac1{6}\left(\Delta\beta\cos^3\theta+3\beta_\bot\cos\theta\right)E^3(t).
\label{eq:hyp_Hamiltonian}
\end{eqnarray}
In these expressions,  
$\theta$ is  the Euler angle between  the internuclear molecular axis and the LFF $Z$-axis,
$\mu$  the permanent electric dipole moment, $\Delta\alpha=\alpha_\parallel-\alpha_\bot$ 
the polarizability anisotropy, with  $\alpha_\bot$ and $\alpha_\parallel$ being its perpendicular 
and parallel components, and $\Delta\beta=\beta_\parallel-3\beta_\bot$  hyperpolarizability anisotropy,
with $\beta_\bot$ and $\beta_\parallel$ being the perpendicular and parallel components, respectively. 
The aim is to explore the dependence of the field-dressed rotational dynamics on the parameters of the two-color laser field,
including its frequency. To do so, the laser frequency is reduced and considered within the regime where the time-average approximation is not 
correct~\cite{Pershan},  but still assuming that the laser electric field 
is non-resonant, \ie, it cannot drive any electronic, vibrational or rotational transition. 

The Hamiltonian~\eqref{eq:molecular_Hamiltonian} is invariant under arbitrary rotations around the 
LFF $Z$-axis $\mathbf{C_Z}(\epsilon)$, and reflections on the LFF $XZ$-plane,
 $\mathbf{\sigma_{XZ}}$. These symmetries imply that the eigenstates associated to  
 Hamiltonian~\eqref{eq:molecular_Hamiltonian}  with a constant electric field are degenerate in 
 $|M|$, with $M$ being the projection of the rotational angular momentum along the LFF $Z$-axis.
The time-dependent Schr\"odinger equation associated to the 
Hamiltonian~\eqref{eq:molecular_Hamiltonian} 
is solved by combining the short iterative Lanczos method~\cite{Beck} for the time variable, 
and a basis set expansion in terms of the field-free basis, formed by
 the spherical harmonics $Y_{J,M}(\Omega)$ with  $\Omega=(\theta,\phi)$ being the Euler angles, 
 including the symmetries of the Hamiltonian, \ie, fixed $M$.
The time-dependent Schr\"odinger equation is solved assuming that at $t=0$ the molecule is in a 
field-free eigenstate, \ie, $\psi(\Omega,t=0)=Y_{J,M}(\Omega)$. 

In this work,   the field-dressed rotational dynamics is analyzed in terms of the expectation values 
\begin{equation}
\expect{\cos^k\theta}=\int\psi^*(\Omega,t)\cos^k\theta\psi(\Omega,t)d\Omega,
\label{eq:expdcg}
\end{equation}
with $\psi(\Omega,t)$ being the time-dependent wave function, and 
positive $k$.
For the orientation  and alignment,  $k=1$ and $k=2$,  respectively. 
The wave function  $\psi(\Omega,t)$  and the expectation values depend on the 
laser field parameters~\eqref{eq:electric_field}, and, therefore, they are invariant under the same 
symmetry transformations as the Hamiltonian~\eqref{eq:molecular_Hamiltonian}.

The two-color electric field~\eqref{eq:electric_field} is invariant under the following
transformation 
\begin{equation}
\mathcal{T}: \left(q_1,q_2,\omega\right) \rightarrow \left(\kappa q_1,\kappa q_2,\frac{\omega}{\kappa}\right) \quad \text{with} \quad \kappa\in\mathbb{Z}^{+}, 
\end{equation}
and,  therefore, the Hamiltonian~\eqref{eq:molecular_Hamiltonian} is also invariant under this transformation. 
As a consequence,  the orientation and alignment are also  invariant under  $\mathcal{T}$. Thus,
the symmetry analysis can be restricted to $q_1$ and $q_2$ satisfying $\text{gcd}(q_1,q_2)=1$.
The symmetries in the phases  $\delta_1$ and $\delta_2$ 
and amplitudes $\epsilon_1$ and $\epsilon_2$
of the two-color electric field~\eqref{eq:electric_field}, imply that 
\begin{eqnarray}
\label{eq:sym_phases}
\expect{\cos^k\theta}(&t&,t_0,\epsilon_1,\epsilon_2,\omega,\delta_1,\delta_2)=\\
\expect{\cos^k\theta}(&t&,t_0,(-1)^{n_1}\epsilon_1,(-1)^{n_2}\epsilon_2,\omega,\delta_1+n_1\pi,\delta_2+n_2\pi), \nonumber
\end{eqnarray}
with $n_1$ and $n_2$ being integers, and $k\in\mathbb{Z}^{+}$. Note that  this expression  shows explicitly  its 
dependence on $t,t_0,\epsilon_1,\epsilon_2,\omega,\delta_1$,  and $\delta_2$ of this expectation value. 
The inversion of the electric field direction gives rise to the following invariance 
\begin{eqnarray}
\label{eq:sym_field}
\expect{\cos^k\theta}(t,t_0,\epsilon_1,\epsilon_2,\omega,\delta_1&,&\delta_2)=\\
(-1)^k\expect{\cos^k\theta}&(&t,t_0,-\epsilon_1,-\epsilon_2,\omega,\delta_1,\delta_2), \nonumber
\end{eqnarray}
with $k\in\mathbb{Z}^{+}$. 
The Hamiltonian~\eqref{eq:molecular_Hamiltonian} is invariant under 
a temporal shift of $t_0$ by changing correspondingly the phases
$\delta_1$ and $\delta_2$, and this expectation value  satisfies 
\begin{eqnarray}
\label{eq:sym_t0}
\expect{\cos^k\theta}(t,t_0,\epsilon_1,\epsilon_2,\omega,\delta_1,\delta_2)&&=\\
\expect{\cos^k\theta}(t,t_0+\tau,\epsilon_1,\epsilon_2,\omega&,&\delta_1-q_1\omega\tau,\delta_2-q_2\omega\tau) \ \ \forall \tau, \nonumber
\end{eqnarray}
$k\in\mathbb{Z}^{+}$. In contrast to the rest of the laser field parameters, the value of $t_0$ cannot be easily 
controlled in an experiment. Therefore, these expectation values are averaged over 
$t_0$~\cite{Casado} as
\begin{equation}
\expect{\expect{\cos^k\theta}}=\frac{\omega}{2\pi}\int_0^{\frac{2\pi}{\omega}}dt_0\expect{\cos^k\theta} \quad \text{with} \quad k\in\mathbb{Z}^{+}, 
\label{eq:t0_average}
\end{equation}
where the integral is restricted to an electric field period due to the  $t_0$ periodicity of the 
electric field~\eqref{eq:electric_field}.

For a one-color electric field, \ie, $\epsilon_1=0$ or $\epsilon_2=0$, the Hamiltonian~\eqref{eq:molecular_Hamiltonian} fulfills the symmetry 
in $t_0$ $H\left(\theta,t_0\right)=H\left(\pi-\theta,t_0+\frac{\pi}{q_i\omega}\right)$, with 
 $\epsilon_i\ne0$ and $i=1$ or $2$,  and the expectation values 
satisfy 
\begin{equation}
\expect{\cos^k\theta}\left(t_0+\frac{\pi}{q_i\omega}\right)=(-1)^k\expect{\cos^k\theta}(t_0),  
\label{eq:sym_one_harm_t0}
\end{equation}
and, therefore
\begin{equation}
\expect{\expect{\cos^k\theta}}=\left[1+(-1)^k\right]\frac{q_i\omega}{2\pi}\int_0^{\frac{\pi}{q_i\omega}}dt_0\expect{\cos^k\theta}, 
\label{eq:sym_one_harm_aver}
\end{equation}
where   the dependence on the field parameters is omitted.
For $k=1$,  relation~\eqref{eq:sym_one_harm_aver} indicates that on average the molecule is not  oriented by 
a one-color laser field,  regardless of its frequency, even if 
all three interactions are taken into account.
Note that for a fixed $t_0$, the molecule is 
oriented, whereas for $t_0+\frac{\pi}{q_i\omega}$, it gets the same orientation but in the opposite direction 
as shown in~\eqref{eq:sym_one_harm_t0}. As a consequence, the $t_0$-averaged orientation becomes zero.
For $k=2$, equation~\eqref{eq:sym_one_harm_aver}  implies the interval of integration in~\eqref{eq:t0_average} can be 
reduced  to $t_0$ to  $0\le t_0\le{\frac{\pi}{q_i\omega}}$.

From the symmetries in~\autoref{eq:sym_phases},  
the $t_0$-averaged expectation value satisfies 
\begin{eqnarray}
\label{eq:sym_t0_d1_d2}
&&\expect{\expect{\cos^k\theta}}(t,\epsilon_1,\epsilon_2,\omega,\delta_1,\delta_2)=\\ 
&&\nonumber \expect{\expect{\cos^k\theta}}(t,(-1)^{n_1}\epsilon_1,(-1)^{n_2}\epsilon_2,\omega,\delta_1+n_1\pi,\delta_2+n_2\pi),
\end{eqnarray}
with $n_1$ and $n_2$ being integers.
The symmetry due to the inversion of the electric field direction~\eqref{eq:sym_field}  reads as
\begin{eqnarray}
 \label{eq:sym_t0_force}
\expect{\expect{\cos^k\theta}}(t,\epsilon_1,\epsilon_2,\omega,\delta_1,\delta_2)&=&\\
(-1)^k\expect{\expect{\cos^k\theta}}(t,-&\epsilon_1&,-\epsilon_2,\omega,\delta_1,\delta_2).
\nonumber
\end{eqnarray}
The invariance on $t_0$~\eqref{eq:sym_t0} gives rise to the following phase-shift symmetry
\begin{eqnarray}
\label{eq:sym_t0_shift}
\expect{\expect{\cos^k\theta}}(t,\epsilon_1,\epsilon_2,\omega,\delta_1,\delta_2)&=&\\
\expect{\expect{\cos^k\theta}}(t,\epsilon_1,&\epsilon_2&,\omega,\delta_1+q_1\Delta,\delta_2+q_2\Delta) 
\nonumber
\end{eqnarray}
with $\Delta$ being an arbitrary phase-shift.

The symmetries~\eqref{eq:sym_t0_d1_d2},~\eqref{eq:sym_t0_force} and~\eqref{eq:sym_t0_shift} imply 
additional  identities for the $t_0$-averaged expectation values.
For $q_1$  and $q_2$ odd integers, it yields
\begin{equation}
\label{eq:sym_t0_parity_1}
\expect{\expect{\cos^{k}\theta}}(t,\epsilon_1,\epsilon_2,\omega,\delta_1,\delta_2)=0,
\end{equation}
with $k$ being an odd integer, which indicates the lack of orientation for $k=1$, and
\begin{eqnarray}
\label{eq:sym_t0_parity_2}
\expect{\expect{\cos^{k}\theta}}(t,\epsilon_1,\epsilon_2,\omega,\delta_1,\delta_2)&=&\\
\expect{\expect{\cos^{k}\theta}}\Big(t,\epsilon_1,\epsilon_2,\omega,\delta_1+n_1\frac{\pi}{2},&
\delta_2&+(2-(-1)^\frac{q_2-q_1}{2})n_1\frac{\pi}{2}\Big),\nonumber  
\end{eqnarray}
with $n_1$ being an integer, and $k$ an even integer. 
Thus, in this case the molecule is aligned but not oriented.
Whereas for $q_1$ odd   and $q_2$ even, it holds
\begin{eqnarray}
\label{eq:sym_d1_d2}
\expect{\expect{\cos^k\theta}}(t,\epsilon_1,\epsilon_2,\omega,\delta_1,\delta_2)&=&\\
(-1)^{k\left(\frac{n_1 q_2}{2}+n_2 q_1\right)}
\expect{\expect{\cos^k\theta}}(t,\epsilon_1,&\epsilon_2&,\omega, 
 \delta_1+n_1\frac{\pi}{2},\delta_2+n_2 \pi) 
\nonumber
\end{eqnarray}
with $n_1$ and $n_2$ being  integers, and,  $k\in\mathbb{Z}^{+}$.  
Hence, in the latter case the molecule is both oriented and aligned.

Due to the symmetries of the Hamiltonian,  the  $t_0$-averaged expectation
values of the orientation and alignment can be expressed as a series expansion in terms of the amplitude and phase,  
\ie, $\epsilon_j$ and $\delta_j$, of the $j$-harmonics due to 
 the two-color electric field~\eqref{eq:electric_field} for fixed time $t$~\cite{Casado}. 
The corresponding expansions~\eqref{eq17A},~\eqref{eq4},~\eqref{fit1}, and~\eqref{fit2}
are derived in the Appendix.

\section{Results}
\label{sec:results}
The carbonyl sulfide molecule OCS serve as prototype for this study.
For OCS~\cite{Maroulis}, the rotational constant is $B=0.20286$~$\text{cm}^{-1}$, the rotational period  
$T_{\text{rot}}=82.2$~ps, the  permanent electric dipole $\mu=0.71$~D, polarizability anisotropy and 
perpendicular term $\Delta\alpha=27.26$~\au, and $\alpha_\bot=26.08$~\au, respectively, hyperpolarizability 
anisotropy and perpendicular term  $\Delta\beta=132.3$~\au, and $\beta_\bot=-59.1$~\au, respectively. 
For $t=0$, the OCS is assumed to be in its rotational ground state, \ie, 
$\psi(\Omega,t=0)=Y_{0,0}(\Omega)$. Although  this work is restricted to the field-dressed rotational dynamics of the 
ground state, similar results are obtained for excited rotational states.

Since the $t_0$-averaged expectation values satisfy the phase-shift symmetry~\eqref{eq:sym_t0_shift}, 
the phase of the first harmonic  is fixed to zero, $\delta_1=0$. 
The electric field strengths are 
taken as $\epsilon_1=(1-\gamma)E_0$ and $\epsilon_2=\gamma E_0$ with $0\le\gamma \le 1$. 
The field strength is related to the laser field intensity as $E_0=\sqrt{\frac{2I}{c\epsilon_0}}$, 
with $c$ being the speed of the light and $\epsilon_0$ the vacuum electric permittivity.   
The laser intensity is fixed to $I=5\cdot 10^{11}$~W/cm$^2$, with the  electric field strength being 
$E_0\approx1.94\cdot 10^{7}$~V/cm. 
Note that this strong laser intensity is routinely used in non-resonant ac laser pulses, 
whereas is larger than the experimentally available intensities of cw lasers~\cite{Deppe:15}. 
However, such a large laser intensity provokes a moderate orientation and a strong 
alignment of the molecule, and these effects can be analyzed  in terms of the field symmetries.
The electric field frequencies are fixed to $q_1=1$ and $q_2=2$, which provoke 
both the orientation and alignment of the molecule.

\subsection{Validity of the time-average approximation}
\label{subsec:time-average}

Despite of not considering a laser pulse, this section is devoted to investigate 
the validity of the time-average approximation~\cite{Pershan}.
For a non-resonant two-color laser field, if the field frequencies, $\omega$ and 
$2\omega$, are far from any molecular resonance and higher than the  molecular rotational frequency, 
the Hamiltonian~\eqref{eq:molecular_Hamiltonian} is  averaged over the rapid oscillations of the non-resonant laser field.
Note that for a laser pulse, it is further assumed that the laser period is much shorter than the pulse duration.
If the time-average approximation would be correct, then the Hamiltonian~\eqref{eq:molecular_Hamiltonian}
would be  reduced to
\begin{eqnarray}
\nonumber 
H&=&B\mathbf{J^2}-\frac1{2}\left(\Delta\alpha\cos^2\theta+\alpha_\bot\right)
f_1(\epsilon_1,\epsilon_2,q_1,q_2,\delta_1,\delta_2)\\
& -&\frac1{6}\left(\Delta\beta\cos^3\theta+3\beta_\bot\cos\theta\right)
f_2(\epsilon_1,\epsilon_2,q_1,q_2,\delta_1,\delta_2)\qquad
\label{eq:H_time_average_1} 
\end{eqnarray}
with 
\begin{eqnarray}
\nonumber 
f_1(\epsilon_1,\epsilon_2,q_1,q_2,\delta_1,\delta_2)&=&\frac{\epsilon_1^2+
\epsilon_2^2}{2}+\epsilon_1\epsilon_2\cos(\delta_1-\delta_2)\delta_{q_1,q_2}\qquad\\
\nonumber 
f_2(\epsilon_1,\epsilon_2,q_1,q_2,\delta_1,\delta_2)&=&
\frac{3}{4}\epsilon_1^2\epsilon_2\delta_{2q_1,q_2}\cos(2\delta_1-\delta_2)\\
\nonumber 
&+&\frac{3}{4}\epsilon_1\epsilon_2^2\delta_{q_1,2q_2}\cos(\delta_1-2\delta_2). 
\end{eqnarray}
For a  two-color electric field with $q_1=1$ and $q_2=2$, the time-averaged  Hamiltonian~\eqref{eq:H_time_average_1} reads
\begin{eqnarray}
\nonumber
H&=&B\mathbf{J^2}-\frac1{4}\left(\Delta\alpha\cos^2\theta+\alpha_\bot\right)(\epsilon_1^2+\epsilon_2^2)\\
&-&\frac1{8}\left(\Delta\beta\cos^3\theta+3\beta_\bot\cos\theta\right)\epsilon_1^2\epsilon_2\cos(2\delta_1-\delta_2). \quad
\label{eq:time-average_Hamiltonian2}
\end{eqnarray}
Thus, depending on the values of the phases $\delta_1$ and $\delta_2$, this time-averaged Hamiltonian
might align the molecule, or both orient and align it.

\begin{figure}[tb]
\centering
\includegraphics[width=0.9\linewidth]{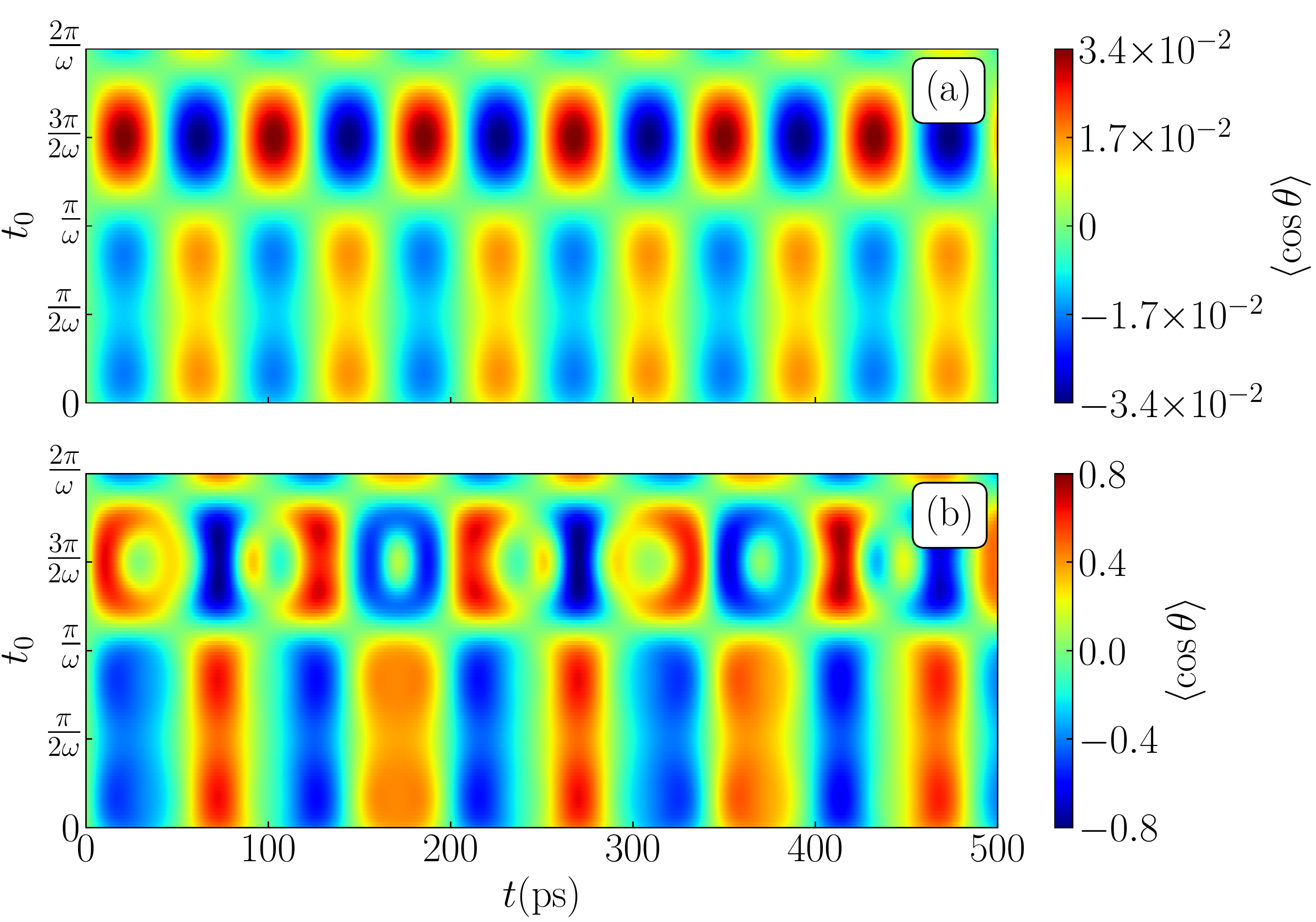}
\caption{For the Hamiltonian $H=H_0+H_\mu$, orientation versus time  and $t_0$
for a two-color laser field with periods (a) $T=10$~fs and (b)  $T=400$~fs. 
The field parameters are fixed to $\gamma=0.5$, $\delta_1=0$, and $\delta_2=\pi/2$.}
\label{fig:validity_time_average_function_of_t0}
\end{figure}

\begin{figure}[tb]
\centering
\includegraphics[width=0.9\linewidth]{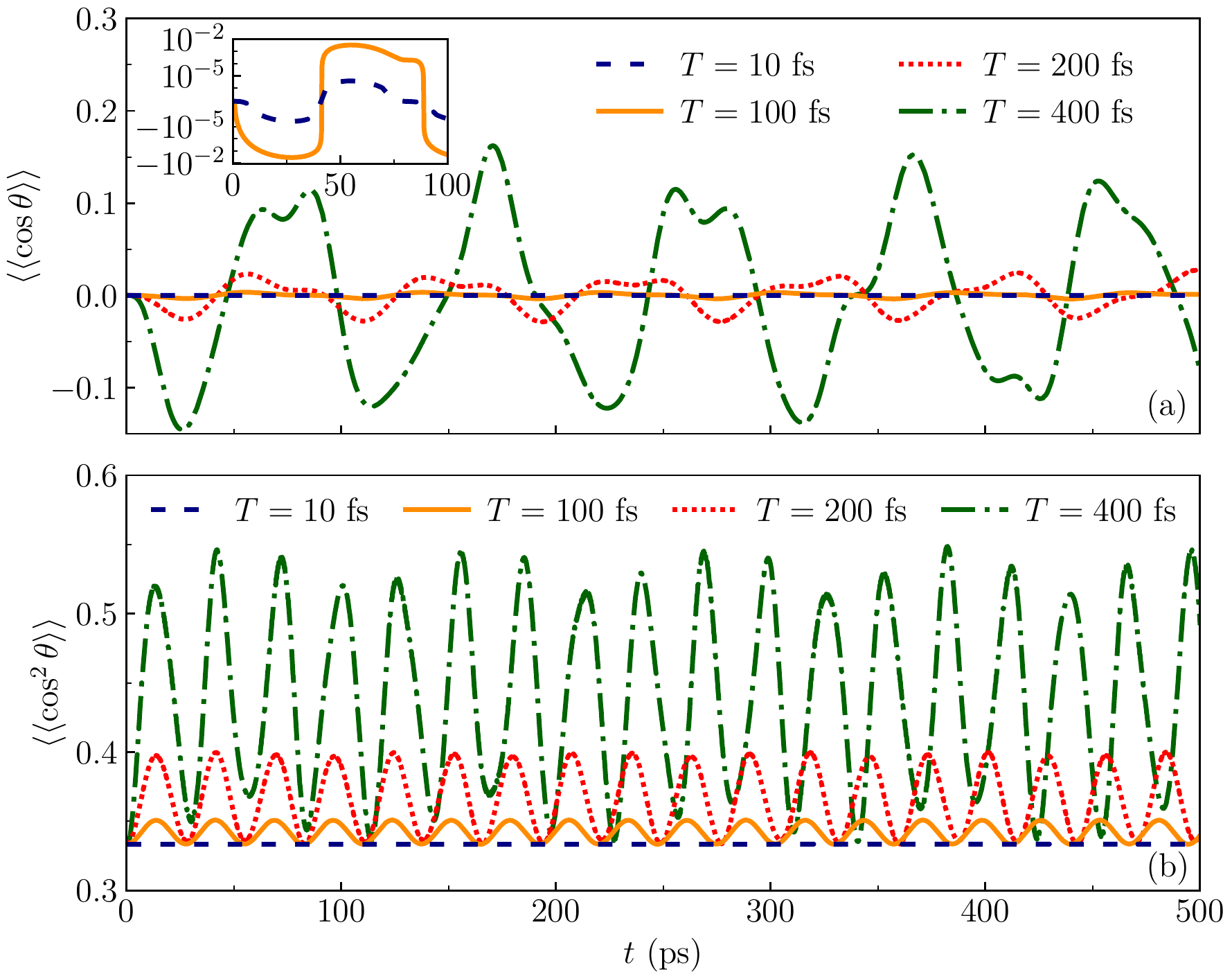}
\caption{For the Hamiltonian $H=H_0+H_\mu$, (a) orientation and (b) alignment averaged over $t_0$ as a function of  time for the 
electric field periods  $T=10$~fs (blue dashed line), $100$~fs (orange solid line), $200$~fs (red dotted line) and 
$400$~fs (green dot-dashed line). The electric field parameters are fixed to $\gamma=0.5$, $\delta_1=0$,
 and $\delta_2=\pi/2$.}
\label{fig:validity_time_average}
\end{figure}

For a one-color laser field, the Hamiltonian~\eqref{eq:time-average_Hamiltonian2} indicates that
if the time-average approximation is correct the molecule should not be oriented.
Regardless of the validity of the time-average approximation, 
 the equality~\eqref{eq:sym_one_harm_aver}  shows that  
 the $t_0$-averaged orientation is zero if only one-color laser field is included in the full Hamiltonian~\eqref{eq:molecular_Hamiltonian}.
 However, by fixing $t_0$ and considering only the interaction with 
 the electric dipole moment, the absolute value of the orientation 
 becomes larger than  $10^{-2}$ for periods larger or equal than $10$~fs and 
$|\expect{\expect{\cos\theta}}|\approx 10^{-8}$;
  whereas for $T=1$~fs,   $|\expect{\cos\theta}|\approx 10^{-3}$ and
 $\expect{\expect{\cos\theta}}\approx 10^{-9}$.
In the regime $T\gtrsim 10$~fs the  time-average  approximation starts to fail. 
The $t_0$-averaged  alignment is non-zero, see~\autoref{eq:sym_one_harm_aver}, 
and the maximal deviation from its field-free value, $\expect{\expect{\cos^2\theta}}=1/3$, is of the order of $10^{-4}$.
Whereas for a fixed $t_0$, the alignment deviation from its field-free value reaches up to 
$10^{-3}$ for a laser period of $10$~fs. If the interactions of the electric field with the polarizability and
hyperpolarizability are also included, the absolute value of the orientation  is also greater than $10^{-2}$ for $T\gtrsim10$~fs and a fixed $t_0$. 
The alignment for a fixed $t_0$ and the $t_0$-averaged alignment reach values up to $0.85$ for a laser period of $10$~fs.

For a two-color laser field, the simplest system including 
only the interaction of the field with the electric dipole moment, \ie, $H=H_0+H_\mu$ is first analyzed.
Figs.~\ref{fig:validity_time_average_function_of_t0}~(a) and (b) show 
 the orientation as a function of $t_0$ and  time for laser field periods $10$~fs and $400$~fs, respectively,
and the field parameters  $\gamma=0.5$ and $\delta_2=\pi/2$.
The orientation is non-zero even for the $10$~fs laser period, and depends on $t_0$. 
These two features contradict the validity of the  time-average approximation.~\autoref{fig:validity_time_average}
presents the $t_0$-averaged orientation and $t_0$-averaged
alignment as a function of the time for the electric field periods $T=10$, $100$, $200$ 
and $400$~fs, and 
$\gamma=0.5$ and $\delta_2=\pi/2$. 
For $T=10$~fs, the $t_0$-averaged orientation is of the order of $10^{-6}$. Note that
 $|\expect{\cos\theta}|$ is  four  orders of magnitude larger, but due to the dependence of 
 $\expect{\cos\theta} $ on $t_0$, see~\autoref{fig:validity_time_average_function_of_t0}~(a),
 the $t_0$-averaged  $\expect{\expect{\cos\theta}}$ becomes very small.
The maximal deviation of the $t_0$-averaged alignment from its field-free value 
is $1.8\times10^{-4}$. Thus, for $T=\frac{2\pi}{\omega}=10$~fs, on average  
the molecule  is not oriented nor aligned. As a consequence, one can mistakenly conclude that  the 
time-average approximation can be applied, however, for a fixed $t_0$, 
it is not correct as shown by the results in~\autoref{fig:validity_time_average_function_of_t0}~(a). 
For $T=100$~fs, the deviations  of $\expect{\expect{\cos\theta}}$ and $\expect{\expect{\cos^2\theta}}$
from the corresponding field-free values are still small but not negligible. 
By increasing the laser field period $T$, \ie, reducing the laser frequency  $\omega$, 
the $t_0$-averaged orientation and $t_0$-averaged alignment increase, see for instance 
$\expect{\expect{\cos\theta}}$ and $\expect{\expect{\cos^2\theta}}$
for $T=200$ and 400~fs.

\begin{figure}[t]
\centering
\includegraphics[width=0.9\linewidth]{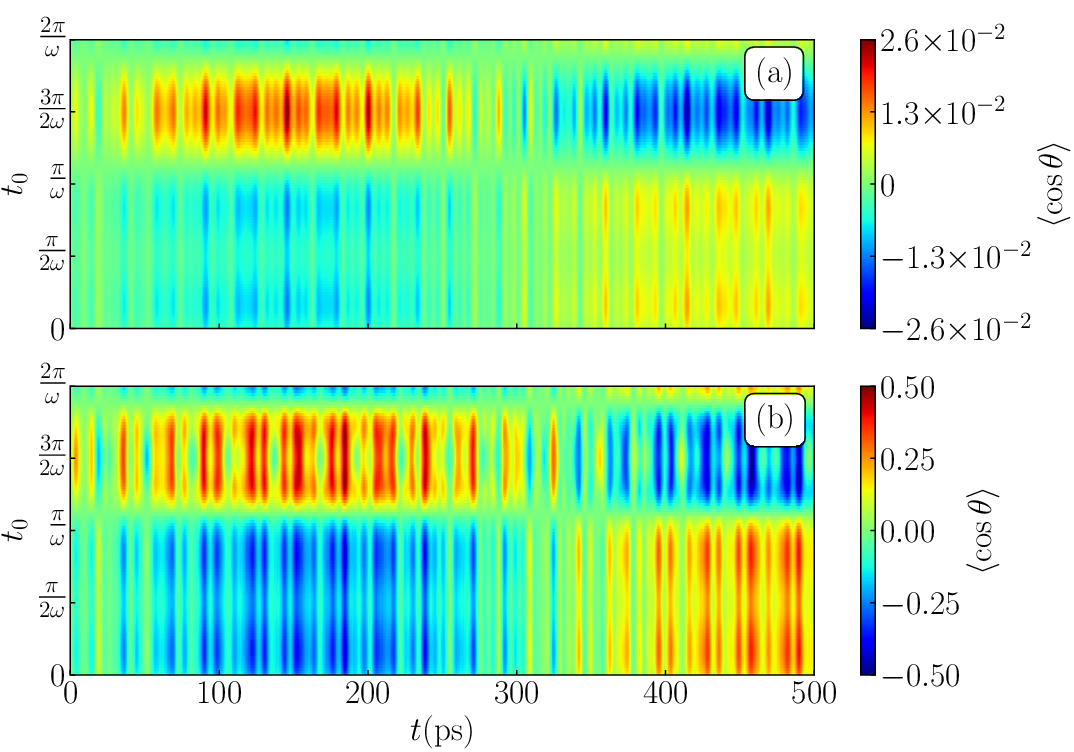}
\caption{For the full  Hamiltonian $H=H_0+H_\mu+H_\alpha+H_\beta$, orientation versus time  and $t_0$
for a two-color laser field with periods (a) $T=10$~fs and (b)  $T=400$~fs. 
The  field parameters are fixed to $\gamma=0.5$, $\delta_1=0$, and $\delta_2=\pi/2$.}
\label{fig:validity_time_average_function_of_t0_All}
\end{figure}

\begin{figure}[t]
\centering
\includegraphics[width=0.9\linewidth]{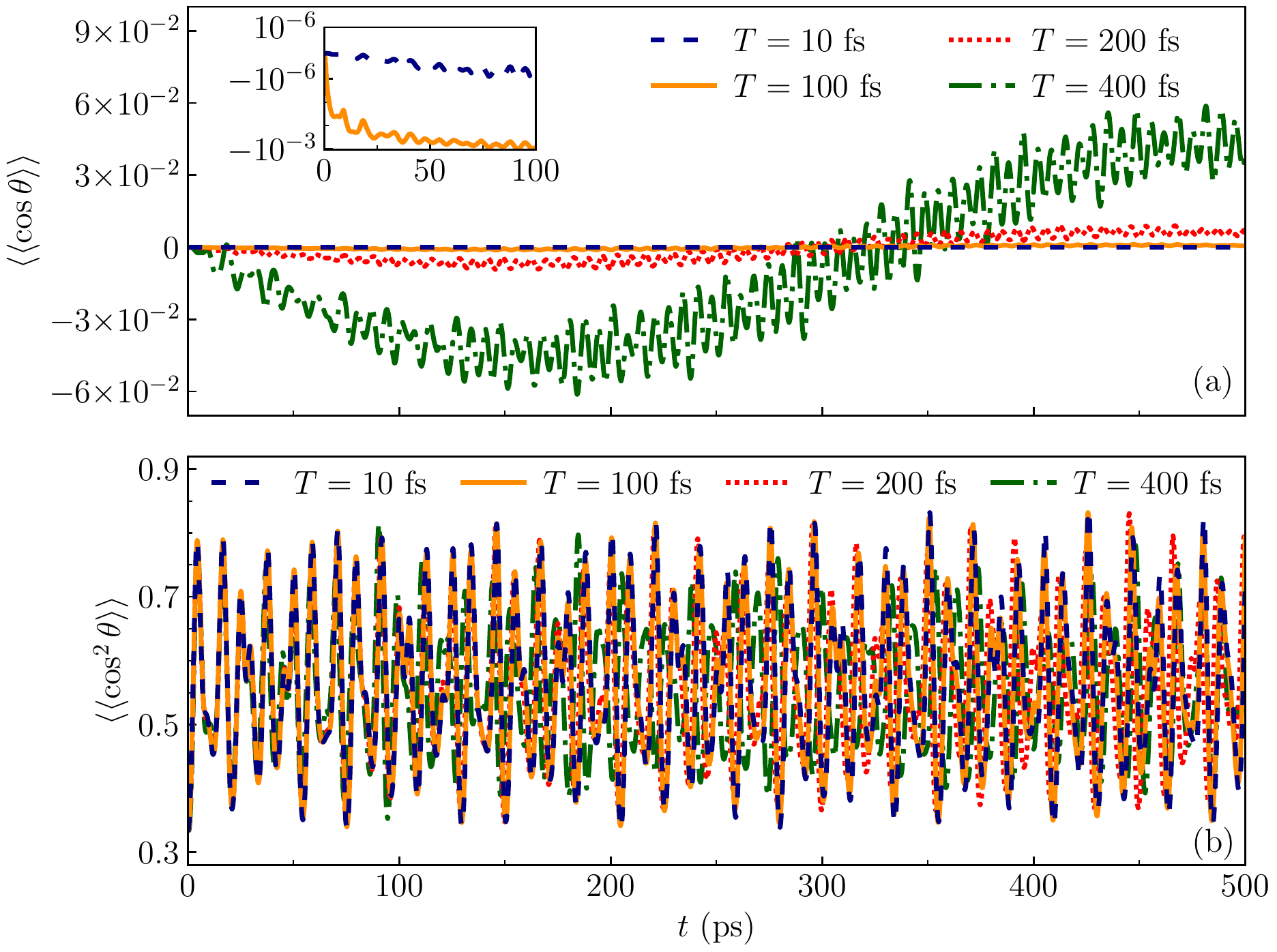}
\caption{For the three interactions, (a) orientation and (b) alignment averaged over $t_0$ versus time for the electric field periods  $T=2\pi/\omega=10$~fs (blue dashed line), $100$~fs (orange solid line), $200$~fs (red dotted line) and 
$400$~fs (green dot-dashed line). The two-color electric field parameters are fixed to $\gamma=0.5$, $\delta_1=0$ and $\delta_2=\pi/2$.}
\label{fig:validity_time_average_All}
\end{figure}

For the  three interactions, \ie, $H=H_0+H_\mu+H_\alpha+H_\beta$, 
Figs.~\ref{fig:validity_time_average_function_of_t0_All}~(a) and (b) show 
 the orientation as a function of $t_0$ and  time for the laser field periods $T=10$~fs and $400$~fs, respectively,
and the electric field parameters  $\gamma=0.5$ and $\delta_2=\pi/2$.
As in the previous case,  $\expect{\cos\theta}$ is non-zero and depends on $t_0$, which indicates that
 even for a  $T= 10$~fs laser, the time-average approximation is not 
 correct.~\autoref{fig:validity_time_average_All} shows the $t_0$-averaged orientation and $t_0$-averaged alignment 
  as a function of the time for the electric field periods $T=10$, $100$, $200$ and $400$~fs. 
 Due to the dependence on $t_0$ of $\expect{\cos\theta}$, the $t_0$-averaged orientation is rather small even for the 
 laser frequency $400$~fs.
 This cancellation effect does not take place when $\expect{\expect{\cos^2\theta}}$ is computed because $\expect{\cos^2\theta}>0$.
 As a consequence, the $t_0$-averaged alignment is very large for all considered laser field periods.

\subsection{Orientation induced by the two-color laser field}
\label{subsec:dip}
In this section, the rotational dynamics is explored, by first focusing on the orientation 
in a  two-color laser  field with period $T=\frac{2\pi}{\omega}=400$~fs  and $\delta_1=0$. 
Note that this electric field period  is two orders of magnitude larger than the period 
of the non-resonant lasers used typically in experiment such as YAG-Laser and Ti-Shaphire~\cite{De:2009}. 
However, such a large electric field period ensures a certain degree of the $t_0$-averaged 
orientation of the molecule, and allows us to 
analyze the field-dressed dynamics in terms of the field symmetries.
This is done by systematically including in the description the interactions of electric field with the 
electric dipole moment, polarizability, and hyperpolarizability of the molecule.

\begin{figure*}[t]
\includegraphics[width=0.97\linewidth]{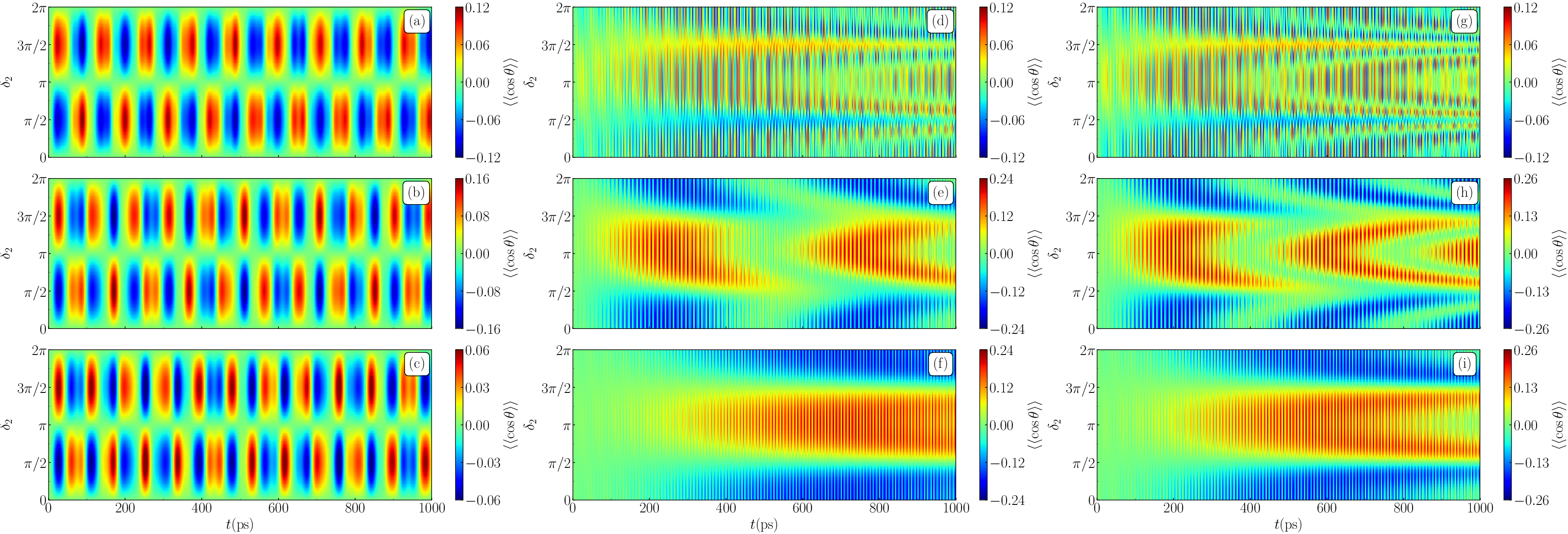}
\caption{Orientation averaged over $t_0$ as a function of the time and the second-harmonic phase for the 
parameters (a), (d) and (g) $\gamma=0.25$, (b), (e), and (h) $\gamma=0.5$ and (c), (f) and (i)
$\gamma=0.75$. The interaction Hamiltonian includes $H_{\mu}$ in panels (a), (b) and (c);  
$H_{\mu}+H_\alpha$ in panels (d), (e) and (f); and $H_{\mu}+H_\alpha+H_\beta$ in panels (g), (h) and (i).}
\label{fig:Orien_g_dip_pola}
\end{figure*}

\begin{figure*}[h]
\centering
\includegraphics[width=0.95\linewidth]{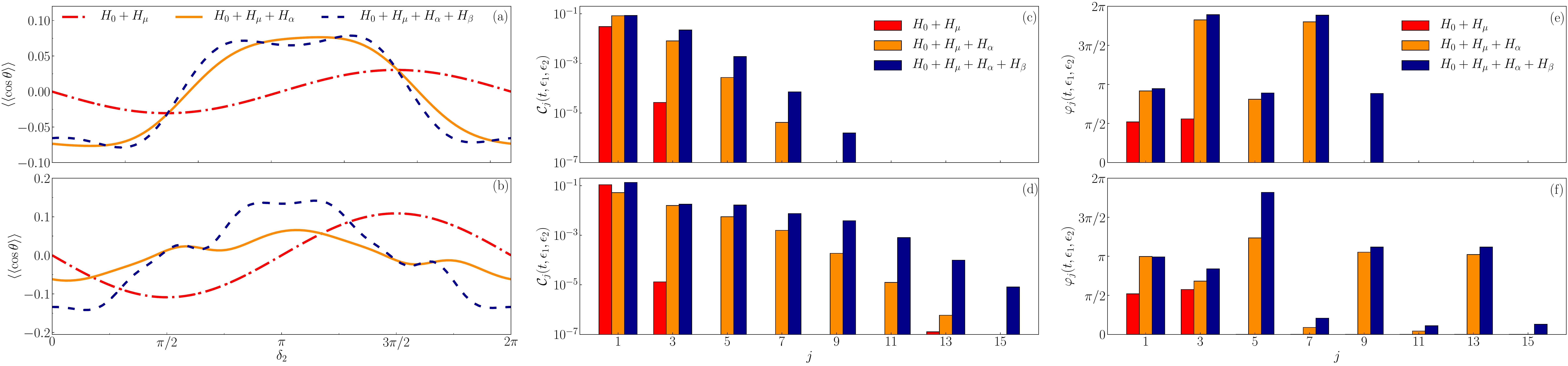}
\caption{For a two-color electric field with $\gamma=0.5$, $t_0$-averaged orientation as a function of 
the second-harmonic phase at fixed propagation times (a) $t=200$~ps and (b) $t=600$~ps.
(c) and (d) fitted coefficients 
${ \cal C}_j(t,\epsilon_1,\epsilon_2)$,   and (e) and (f) fitted phases  
$\varphi_j(t,\epsilon_1,\epsilon_2)$ from  the analytic expression~\eqref{fit1} of 
$\expect{\expect{{\cos\theta}}}$ for $t=200$~ps and $t=600$~ps, respectively.  The  interaction Hamiltonian includes 
$H_{\mu}$ (red dot-dashed line), $H_{\mu}+H_\alpha$ (orange solid line), and $H_{\mu}+H_\alpha+H_\beta$ (blue dashed line).}
\label{fig:Orien_g_05_t_fixed}
\end{figure*}

By considering only the interaction of the electric field with its 
permanent electric dipole moment, \ie, $H=H_0+H_{\mu}$, the contour 
plots~\autoref{fig:Orien_g_dip_pola}~(a),~(b) and~(c) present the $t_0$-averaged orientation as a function 
of the propagation time $t$ and the phase of the second-harmonic $\delta_2$ for the strength parameters
$\gamma=0.25$, $0.5$ and $0.75$, respectively.
For a fixed time, $\expect{\expect{\cos\theta}}$ 
satisfies the symmetry~\eqref{eq:sym_d1_d2} for $k=1$, $n_1=0$,  and $n_2=1$,
and 
approximately fulfills the relation 
$\expect{\expect{\cos\theta}}(t,\delta_2)\approx\expect{\expect{\cos\theta}}(t,\pi-\delta_2)$.
Regardless of the values of $\gamma$ and $t$,  $\expect{\expect{\cos\theta}}$ shows the same
dependence on $\delta_2$, reaching its maximal orientation in absolute value for 
$\delta_2\approx\pi/2$ and $\delta_2\approx3\pi/2$, and the minimal one for $\delta_2\approx0$ and 
$\delta_2\approx\pi$.
For a certain $\gamma$ and $\delta_2$,  $\expect{\expect{\cos\theta}}$ oscillates as a function of time, and 
the field-dressed  wave function has contributions of only few field-free states. 
Note that for $\delta_2\approx0,\pi$, the amplitude of these oscillations is very small.

By adding the interaction of the electric field with the molecular polarizability, \ie, $H=H_0+H_{\mu}+H_{\alpha}$, the field-dressed 
dynamics becomes more complex. The corresponding $t_0$-averaged orientation is presented 
in~\autoref{fig:Orien_g_dip_pola}~(d),~(e) and~(f) for $\gamma=0.25$, $0.5$ and $0.75$,  respectively.
In this case, $\expect{\expect{\cos\theta}}$ also satisfies the symmetry relation \eqref{eq:sym_d1_d2}  for $k=1$, $n_1=0$,  and $n_2=1$. 
In addition, the dependence of  $\expect{\expect{\cos\theta}}$ on the second-harmonic phase for fixed time $t$ strongly 
depends on the parameter $\gamma$, \ie, on the relative weight of the electric field components. 
For $\gamma=0.25$, the $t_0$-averaged orientation is composed of slow oscillations with superimposed fast 
modulations of the amplitude, and it is lower than $0.12$.
In contrast, an orientation up to $0.24$ is achieved for $\gamma=0.5$ and $\gamma=0.75$, and $\expect{\expect{\cos\theta}}$
slowly oscillates with time, whereas the amplitude also show small oscillations.

The $t_0$-averaged orientation when the three interactions are considered, \ie, 
$H=H_0+H_{\mu}+H_{\alpha}+H_{\beta}$,  is presented in~\autoref{fig:Orien_g_dip_pola}~(g),~(h) and~(i) 
for $\gamma=0.25$, $0.5$ and $0.75$, respectively. The rotational dynamics in this case shows a qualitatively similar 
behavior as when $H_\beta$ is neglected, 
compare panels (d)-(g), (e)-(h) and (f)-(i) of~\autoref{fig:Orien_g_dip_pola}. The absolute value of the 
orientation is slightly larger in this case, and for a fixed $\delta_2$, the oscillations as  a function of $t$ show smaller periods.

For a fixed configuration of the two-color laser field and a certain propagation time $t$,  the 
analytic expression~\eqref{fit1}  of $\expect{\expect{\cos\theta}}$, and its dependence 
on the phase of the second harmonics $\delta_2$  is illustrated in~\autoref{fig:Orien_g_05_t_fixed}.
Panels ~(a) and~(b) in~\autoref{fig:Orien_g_05_t_fixed}
show the $t_0$-averaged orientation as a function $\delta_2$ for propagation times $t=200$~ps and 
$t=600$~ps, respectively, and the two components of the electric field having the same weight, \ie, 
$\gamma=0.5$. These curves have been numerically fitted to the expansion~\eqref{fit1} 
using the $\delta_2$-independent constants
${ \cal C}_j(t,\epsilon_1,\epsilon_2)$ and 
$\varphi_j(t,\epsilon_1,\epsilon_2)$ as fitting parameters with $j$ being an odd integer. 
Figs.~\ref{fig:Orien_g_05_t_fixed}~(c) and~(d) show these fitted coefficients 
${ \cal C}_j(t,\epsilon_1,\epsilon_2)$ at $t=200$~ps and $t=600$~ps, 
respectively, the fitted phases are shown in Figs.~\ref{fig:Orien_g_05_t_fixed}~(e) and~(f). If only the 
interaction with the electric dipole moment is included, the first coefficient 
$j=1$ is sufficient to reproduce 
the dependence of $\expect{\expect{\cos\theta}}$ in $\delta_2$ with a fairly good accuracy, and the phase 
of this $j=1$ coefficient is close to $\pi/2$, in agreement with the observed sine-like behaviour,
\ie, $\expect{\expect{\cos\theta}}(t,\delta_2)\approx\expect{\expect{\cos\theta}}(t,\pi-\delta_2)$. Note that 
the next term in the expansion~\eqref{fit1} with $j=3$ is smaller than $5\times10^{-5}$ for these two 
propagation times. The deviation of the phases $\varphi_1(t,\epsilon_1,\epsilon_2)$ and  $\varphi_3(t,\epsilon_1,\epsilon_2)$ from being exactly $\pi/2$ prevents the  $t_0$-averaged orientation 
from being exactly  zero at $\delta_2=0$ and $\pi$. 
By adding the interaction with the molecular polarizability, higher 
order terms becomes more important in~\autoref{fit1}. Indeed, the 
 $j\ge 7$ ($j\ge 11$) coefficients are smaller than  
$10^{-4}$ for $t=200$~ps ($t=600$~ps). Finally, if the three interactions are considered, \ie, 
$H=H_0+H_{\mu}+H_\alpha+H_\beta$, the dependence of $\expect{\expect{\cos\theta}}$ in $\delta_2$ 
gets more complicated, and the contribution of higher order terms increases  gaining importance on the 
expansion~\eqref{fit1}. In these two cases, the fitted phases take values very close to 
$\pi$ or $2\pi$, see Figs.~\ref{fig:Orien_g_05_t_fixed}~(e) and~(f), and the deviation from these values prevents $\expect{\expect{\cos\theta}}$ from being 
zero at $\delta_2=\pi/2$ and $3\pi/2$.

\begin{figure*}[t]
\includegraphics[width=0.97\linewidth]{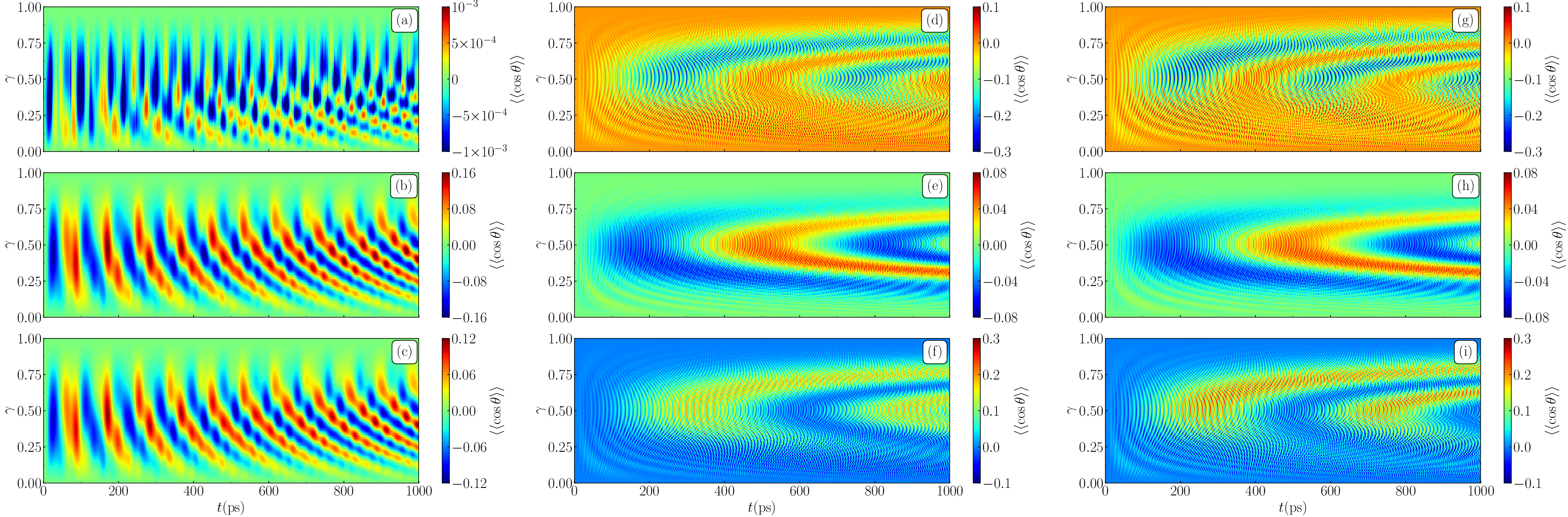}
\caption{The $t_0$-averaged orientation as a function of the propagation time and of the relative weight of the two electric field 
components $\gamma$ for the phase of the second-harmonic (a), (d), and (g) $\delta_2=0$; (b), (e) and (h) $\delta_2=\pi/2$;  and 
(c), (f) and (i) $\delta_2=3\pi/4$. 
The interaction Hamiltonian includes  $H_{\mu}$ in panels (a), (b) and (c);   $H_{\mu}+H_\alpha$ in panels  (d), (e) and (f); 
and $H_{\mu}+H_\alpha+H_\beta$ in panels (g), (h) and (i).}
\label{fig:Orien_d_dip}lo
\end{figure*}

This analysis is completed by investigating the dependence of the $t_0$-averaged orientation on the
relative weight of the two electric field components $\gamma$ in~\autoref{fig:Orien_d_dip}. When only 
the interaction with the electric dipole moment is taken into account, the results for $\delta_2=0$, $\pi/2$, 
and $3\pi/4$ are presented in~\autoref{fig:Orien_d_dip}~(a),~(b) and~(c), respectively. As discussed 
in~\autoref{subsec:time-average}, the $t_0$-averaged orientation is zero for $\gamma=0$ and 
$\gamma=1$ because the electric field~\eqref{eq:electric_field} becomes one-color. For fixed $\gamma$ 
and $\delta_2$, $\expect{\expect{\cos\theta}}$ shows fast oscillation versus $t$. The dependence of 
$\expect{\expect{\cos\theta}}$ on $\gamma$ changes as the propagation time $t$ increases. The 
orientation tends to reach larger values for $0.25\lesssim\gamma\lesssim0.75$.
The maximal orientation  is $0.16$, which is achieved for $\delta_2=\pi/2$.
The orientation for $\delta_2=0$ is non zero but  lower than $10^{-3}$. 

For $H=H_0+H_{\mu}+H_\alpha$,  Figs.~\ref{fig:Orien_d_dip}~(d),~(e) and~(f) present  
$\expect{\expect{\cos\theta}}$ for $\delta_2=0$, $\pi/2$ and $3\pi/4$,  respectively. Compared to the 
previous case, the dependence of $\expect{\expect{\cos\theta}}$ on $\gamma$ and $t$ is significantly 
changed. As a function of $t$, $\expect{\expect{\cos\theta}}$ shows slow oscillations, whose
amplitude is modulated.
A significant orientation of the molecule is  attained for several values of $\gamma$.
 The maximal 
absolute value of the orientation is reached for $\delta_2=0$, $\pi$ (not shown here), 
and  $3\pi/4$, whereas the minimal one  occurs for $\delta_2=\pi/2$. 
Indeed, for $\delta_2=\pi/2$, $|\expect{\expect{\cos\theta}}|$ is smaller than $0.08$, and it 
reaches close to $0.3$ for $\delta_2=3\pi/4$ and $0$, respectively.
 By adding the 
interaction with the hyperpolarizability, \ie, $H=H_0+H_{\mu}+H_\alpha+H_\beta$, the $t_0$-averaged 
orientation is not significantly modified, see~\autoref{fig:Orien_d_dip}~(g),~(h) and~(i). In particular,
$\expect{\expect{\cos\theta}}$ shows a qualitatively similar dependence on $t$ and $\gamma$ as in the 
previous case.

\subsection{Alignment induced by the two-color laser field}
\label{subsec:alig}
\begin{figure}
\centering
\includegraphics[width=0.9\linewidth]{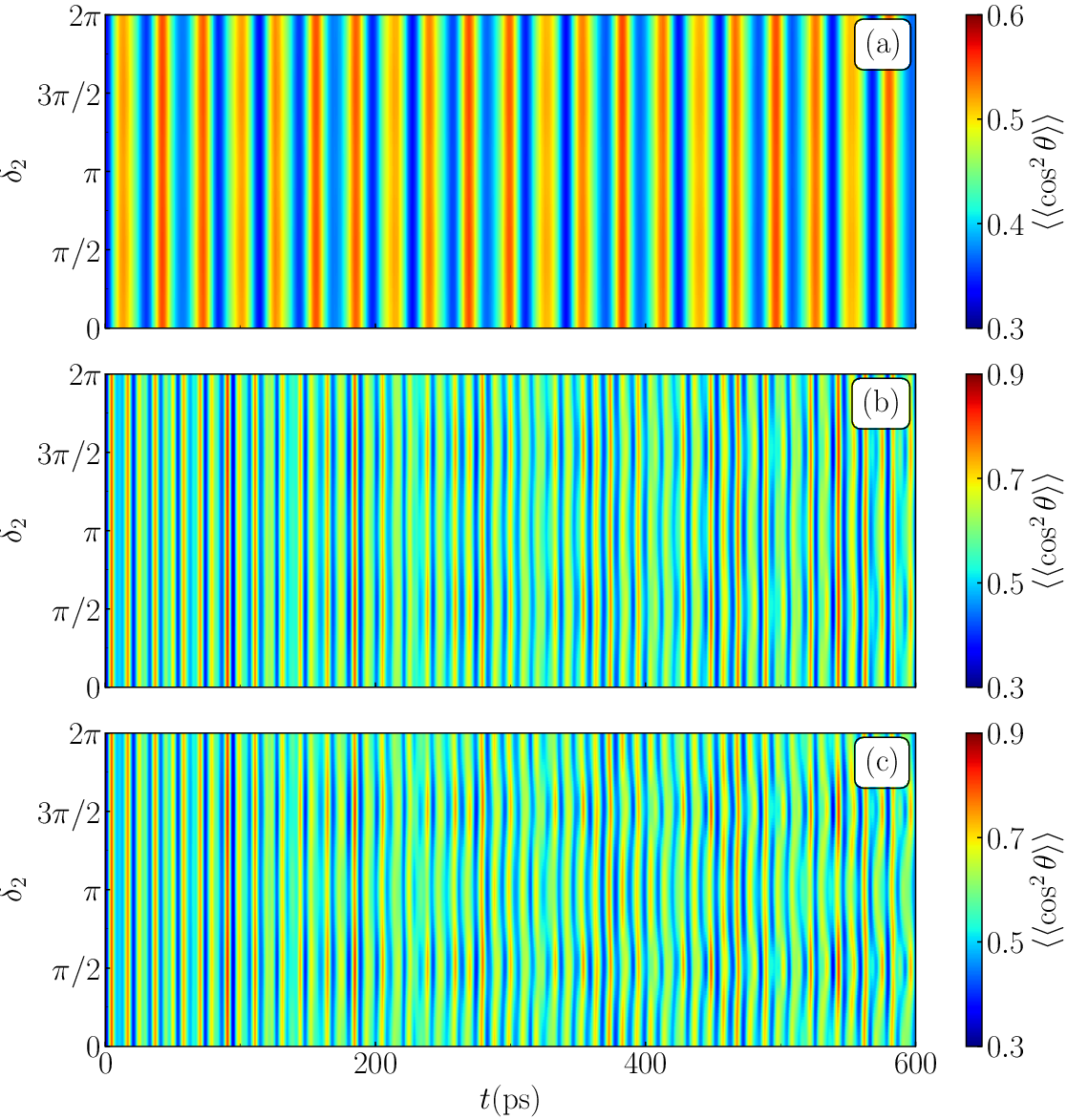}
\caption{The $t_0$-averaged alignment a function of the propagation time and the second-harmonic phase $\delta_2$ for the relative 
weight of  the electric field components  $\gamma=0.5$. The interaction Hamiltonian includes (a) $H_\mu$, (b) $H_\mu+H_\alpha$, and 
(c) $H_\mu+H_\alpha+H_\beta$.} \label{fig:Alig_g_all}
\end{figure}

This section presents  the $t_0$-averaged alignment of the molecule in a  two-color laser field with period 
$T=400$~fs.   
The panels (a), (b) and (c) of \autoref{fig:Alig_g_all} present $\expect{\expect{\cos^2\theta}}$ versus 
$\delta_2$ and $t$ when the two electric field components have the same weight $\gamma=0.5$ and  by 
including progressively the three interactions with the two-color electric field in the Hamiltonian. 
For all considered configurations, the $t_0$-averaged
alignment satisfies the symmetry relation~\eqref{eq:sym_d1_d2} for $k=2$, $n_1=0$ and $n_2=1$.
For $H=H_0+H_{\mu}$,  $\expect{\expect{\cos^2\theta}}$ depends very weakly on $\delta_2$, and oscillates 
as $t$ increases quasi-periodically between $0.3$ and $0.6$, see~\autoref{fig:Alig_g_all}~(a).
By also taking into account the interaction with the molecular polarizability and 
with the hyperpolarizability, $\expect{\expect{\cos^2\theta}}$ show a rather weak dependence on $\delta_2$ 
for short propagation times, which becomes stronger for $t\gtrsim200$~ps, see 
Figs.~\ref{fig:Alig_g_all}~(b) and~(c). 
In these cases, the oscillations of $\expect{\expect{\cos^2\theta}}$ versus $t$ are faster and reach 
up to $0.9$, indicating that the molecule is strongly aligned.

\begin{figure*}
\centering
\includegraphics[width=0.95\linewidth]{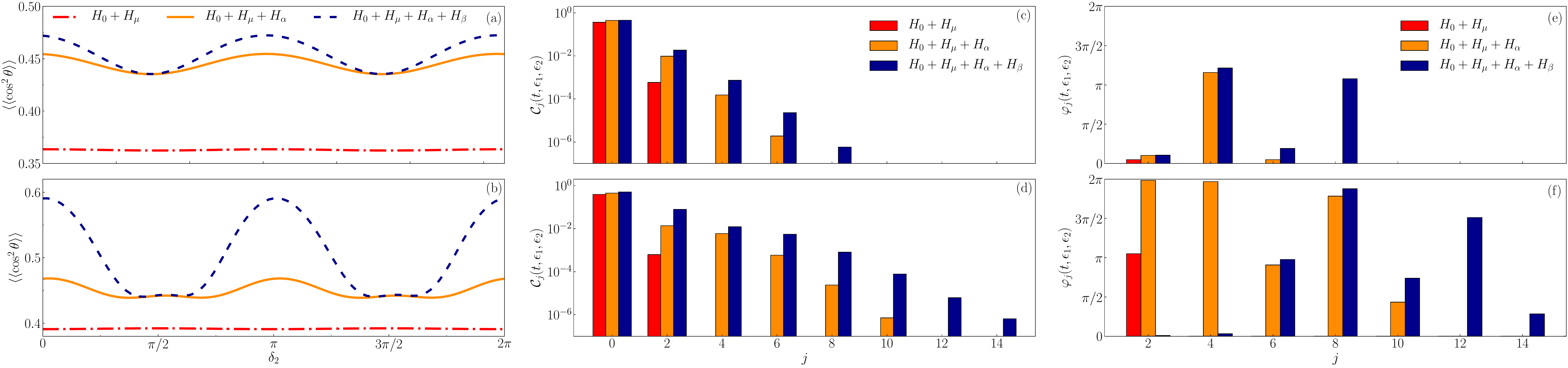}
\caption{For a two-color electric field with $\gamma=0.5$, $t_0$-averaged alignment as a function of 
the second-harmonic phase at fixed propagation times (a) $t=200$~ps and (b) $t=600$~ps. 
(c) and (d) fitted coefficients ${ \cal C}_j(t,\epsilon_1,\epsilon_2)$,   and (e) and (f) fitted phases  
$\varphi_j(t,\epsilon_1\,\epsilon_2)$ from  the analytic expression~\eqref{fit2} of 
$\expect{\expect{{\cos^2\theta}}}$ for $t=200$~ps and $t=600$~ps, respectively. The interaction Hamiltonian includes $H_{\mu}$ 
(red dot-dashed line), $H_{\mu}+H_\alpha$ (orange solid line),  and $H_{\mu}+H_\alpha+H_\beta$ (blue dashed line).}
\label{fig:Align_g_05_t_fixed}
\end{figure*}

The analytic expression~\eqref{fit2} provides the dependence on
$\delta_2$ of the expectation values  $\expect{\expect{\cos^2\theta}}$. In~\autoref{fig:Align_g_05_t_fixed}, 
$\expect{\expect{\cos^2\theta}}$ is plotted as a function of the second-harmonic phase for $\gamma=0.5$ 
and propagation times $t=200$~ps and $t=600$~ps. The fitted 
coefficients ${ \cal C}_j(t,\epsilon_1,\epsilon_2)$ and phases  $\varphi_j(t,\epsilon_1,\epsilon_2)$, with 
$j$ being an even integer, of these numerical results 
to the series~\eqref{fit2} are presented in~\autoref{fig:Align_g_05_t_fixed}~(c-d), and~(e-f), 
respectively.
When only the  electric field interaction with the electric dipole moment is taken into account, the alignment 
does not depend on $\delta_2$, and the $j=0$ coefficient is enough to accurately reproduce this result, 
\ie, ${ \cal C}_0(t,\epsilon_1,\epsilon_2)$ is  only the fitting  parameter.
The next term ${ \cal C}_2(t,\epsilon_1,\epsilon_2)$ is smaller than 
$2\times 10^{-4}$.
For $H=H_0+H_\mu+H_\alpha$, higher order terms are needed in the 
$\expect{\expect{\cos^2\theta}}$ analytical expansion, and they become smaller than $10^{-4}$ for 
$j\ge 6$ and $j\ge 8$ for $t=200$~ps and $t=600$~ps, respectively.
By adding $H_\beta$, \ie, $H=H_0+H_\mu+H_\alpha+H_\beta$, the $\delta_2$-dependence of 
$\expect{\expect{\cos^2\theta}}$ becomes more complex, and even higher order terms are 
required for an accurate fitting. 
For the fitting phases $\varphi_j(t,\epsilon_1,\epsilon_2)$
in these two cases, a broad range of values is encountered, see Figs.~\ref{fig:Align_g_05_t_fixed}~(e) and (f).

\begin{figure}
\centering
\includegraphics[width=0.9\linewidth]{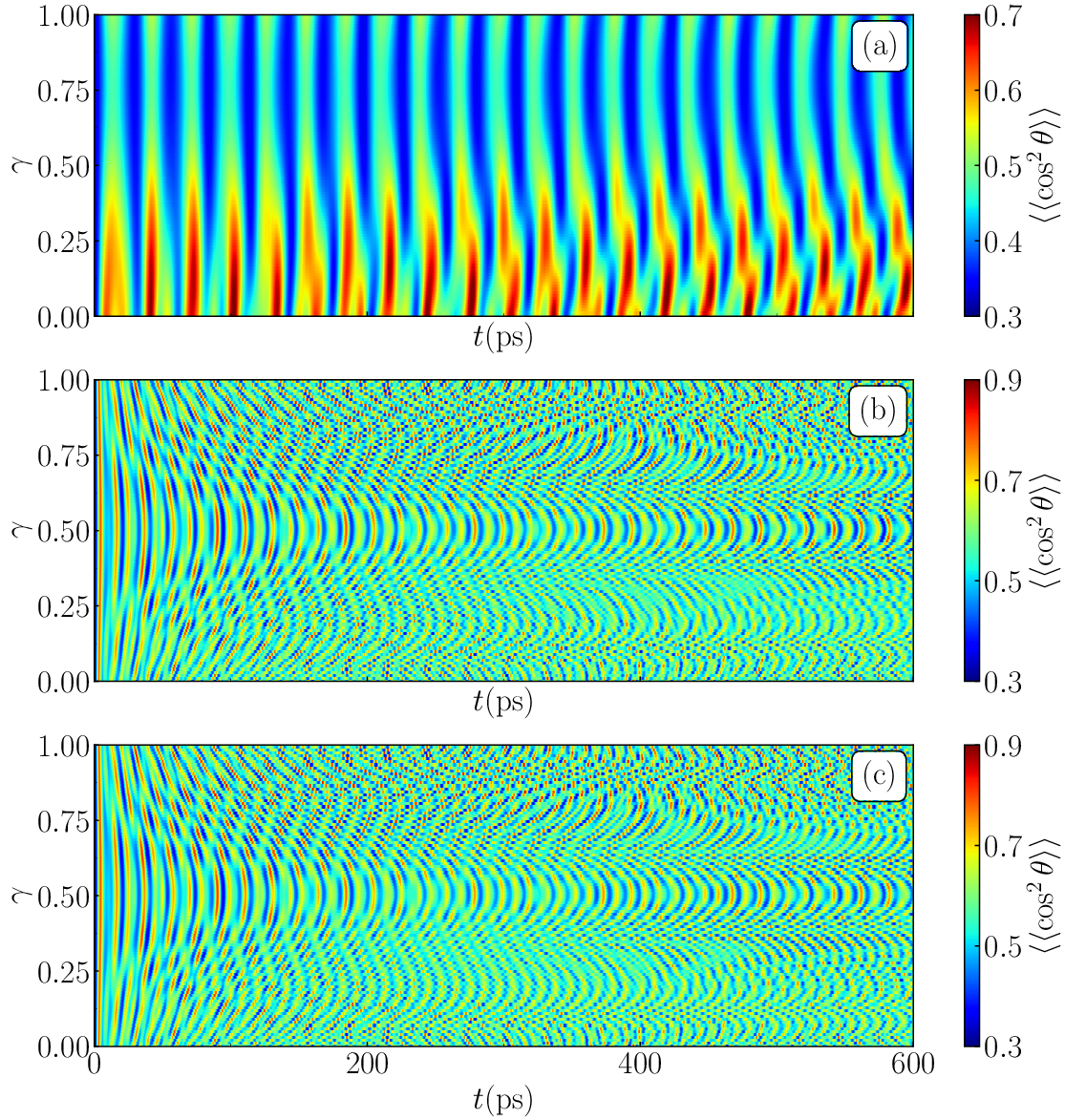}
\caption{The $t_0$-averaged alignment as a function of the propagation time and the relative weight of the 
electric field components $\gamma$ for the second-harmonic phase $\delta_2=3\pi/4$. The 
interaction Hamiltonian includes (a) $H_\mu$, (b) $H_\mu+H_\alpha$, and (c) $H_\mu+H_\alpha+H_\beta$.}
\label{fig:Alig_d_all}
\end{figure}

To conclude, the dependence of $\expect{\expect{\cos^2\theta}}$  on $\gamma$ and $t$ is illustrated 
in~\autoref{fig:Alig_d_all} for 
$\delta_2=3\pi/4$. If only the electric field interaction with the electric 
dipole moment is taken into account, see~\autoref{fig:Alig_d_all}~(a), the $t_0$-alignment oscillates 
between $0.3$ and $0.7$, and reaches the largest values in the region $\gamma<0.5$, \ie, when the field
strength of the first-harmonic is bigger than the one of the second-harmonic. 
For fixed $\gamma$, $\expect{\expect{\cos^2\theta}}$ shows quite regular oscillations versus time.
By taking into account the electric field interaction with the polarizability and with both polarizability and 
hyperpolarizability,~\autoref{fig:Alig_d_all}~(b) and~\autoref{fig:Alig_d_all}~(c) respectively, 
$\expect{\expect{\cos^2\theta}}$ reaches larger values, up to $0.9$, and the frequency of its  oscillations is increased.

\section{Conclusions}
\label{sec:conclusions}

The impact of a two-color continuous-wave non-resonant laser field in the rotational 
dynamics of a linear polar molecule has been investigated. Working within the Born-Oppenheimer and the rigid-rotor 
approximations, the symmetries of the Hamiltonian have been explored and their effects on the field-dressed 
rotational dynamics. 
Due to these symmetries, it is not possible to orient the polar molecule  with a one-color 
cw laser field being necessary to employ a two-color one having odd and even products of the laser 
frequency.

The validity of the time-average 
approximation is first investigated assuming that this cw-non-resonant field does not drive any electronic, vibrational or rotational transitions, 
and despite that the laser field is  not a pulse as normally used in experiments. 
For a cw-laser, this approximation starts to fail for laser periods satisfying $T\gtrsim 10$~fs. 
The field-dressed rotational dynamics has been analyzed in the regime where the time-average 
approximation is not correct. By systematically including the interaction of the field with the electric dipole moment, 
polarizability and hyperpolarizability in the Hamiltonian, the $t_0$-averaged orientation and alignment have been 
analyzed versus the two-color laser field phase and relative strength of the two components. If only the 
interaction with the electric dipole moment is taken into account, the orientation has sinusoidal dependence 
on this relative phase, whereas the alignment is independent of it. Being possible to reproduce these numerical 
results with single-term analytical expressions. By considering the interaction with the molecular 
polarizability, and, in addition with the hyperpolarizability, the orientation and alignment show a more 
complex, but still symmetric, dependence on the phase. As a consequence, more terms on the analytical 
expression are required to properly reproduce the numerical results. Regarding their dependence on the 
relative weight of the two electric field components, the largest $t_0$-averaged orientation and alignment 
are not necessarily obtained when both have the same weight.

Although this study is restricted to the OCS molecule, the observed physical phenomena 
occur in other polar molecules. A natural extension to this work would be to consider a
linear molecule in a two-color cw laser fields with  the two electric field components having perpendicular 
polarizations~\cite{mun}. In this field-configuration,  the symmetries of the system are reduced, and the 
two field components tend to align and orient the molecule in different directions. 
In addition, more  complex molecules, such as symmetric or asymmetric tops, in 
two-color cw non-resonant laser field could be also explored.

\begin{acknowledgments}  
N.R.Q. acknowledges the financial support from the Alexander von Humboldt Foundation. 
Financial support by the Spanish Project No. FIS2017-89349-P  (MINECO), and by the Andalusian 
research  group FQM-207 is  gratefully acknowledged. 
This study has been partially  financed by the Consejer\'{\i}a de 
Conocimiento, Investigaci\'on y Universidad, Junta de  Andaluc\'{\i}a and European Regional Development 
Fund (ERDF), Ref. SOMM17/6105/UGR. 
\end{acknowledgments}

\appendix
\section{Analytic expressions of the orientation and alignment}
\label{sec:appendix}

Following the results of Ref.~\cite{quintero:2010,Cuesta,Casado},   
the orientation and the alignment can be expressed in terms of the amplitude and  phase,  $\epsilon_j$ and $\delta_j$,  of the $j$-harmonics
$g_j(t,t_0,\epsilon_{j},\delta_{j})=\epsilon_{j}\,\cos[q_j \omega (t+t_0)+\delta_j]$, with
$j=1,\dots,s$, appearing in the  Hamiltonian~\eqref{eq:molecular_Hamiltonian} due to the 
interaction of the molecule with the two-color electric field~\eqref{eq:electric_field}. 
The frequencies, amplitudes, and phases of these harmonics are collected in~\autoref{tab1}. 
The first two rows in~\autoref{tab1}  with $j=1$ and $2$ provide the harmonics of the biharmonic field, which appear 
due to the interaction of the biharmonic electric field 
with the permanent electric dipole moment.
By including also the interaction of the biharmonic electric field  with the molecular polarizability, a term proportional to
$E^2(t)$ appears,  and the harmonics 
with $j=3,\dots,6$, see~\autoref{tab1}, also contribute to the Hamiltonian. The cubic term $E^3(t)$ is due to interaction with 
the molecular hyperpolarizability, and is responsible for the 
harmonics with $7\le j\le 14$ in~\autoref{tab1}.
\begin{table}
	\begin{tabular}{|c|c|c|c|}
		\hline
		$j$    &    $q_j$    &   $\epsilon_j$       &    $\delta_j$ \\
		\hline 
		\hline 

		1      &    $q_1$      &   $\epsilon_1$       &    $\delta_1$ \\
		\hline 
		2      &    $q_2$      &   $\epsilon_2$       &    $\delta_2$ \\
		\hline 
		3      &    $2q_1$      &   $\epsilon_1^2/2$   &    $2 \delta_1$ \\
		\hline
		4      &    $2q_2$      &   $\epsilon_2^2/2$   &    $2 \delta_2$ \\
		\hline	
		5      &    $q_1+q_2$      &   $\epsilon_1 \epsilon_2$   &    $\delta_1+\delta_2$ \\	
		\hline
		6      &    $q_2-q_1$      &   $\epsilon_1 \epsilon_2$   &    $\delta_2-\delta_1$ \\
		\hline
		7      &    $q_1$      &    $(3/2) \epsilon_1 \epsilon_2^2+(3/4)  
		\epsilon_1^3$      &    $\delta_1$ \\
		 \hline
		 8      &    $q_2$      &   $(3/2) \epsilon_1^2 \epsilon_2 +(3/4)
		 \epsilon_2^3$     &    $\delta_2$ \\
		\hline 
		9      &    $3q_1$      &   $\epsilon_1^3/4$   &    $3 \delta_1$ \\
		\hline
		10     &    $3q_2$      &   $\epsilon_2^3/4$   &    $3 \delta_2$ \\
		\hline
		11      &    $q_2+2q_1$      &   $(3/4) \epsilon_1^2 \epsilon_2$   &    $\delta_2+2\delta_1$ \\
		\hline
		12      &    $q_2-2q_1$      &   $(3/4) \epsilon_1^2 \epsilon_2$   &    $ \delta_2-2\delta_1$ \\
		\hline
		13      &    $2q_2+q_1$      &   $(3/4) \epsilon_1 \epsilon_2^2$   &    $2\delta_2+\delta_1$ \\
		\hline
		14      &    $2q_2-q_1$      &   $(3/4) \epsilon_1 \epsilon_2^2$   &    $2 \delta_2 -\delta_1$ \\
		\hline	
	\end{tabular}
	\caption{Prefactor of the frequency $\omega$, $q_j$, amplitude, $\epsilon_j$ and phase, $\delta_j$,  
	 of the $j$-harmonics	appearing in the Hamiltonian~\eqref{eq:molecular_Hamiltonian}
	  due to the interaction of the molecule with the two-color electric field~\eqref{eq:electric_field}.}
	\label{tab1}
\end{table}

%

Based on simple symmetry considerations of the $s$ harmonic functions, 
see Ref.~\cite{Cuesta,Casado},  
the orientation and the alignment can be expressed  as
\begin{eqnarray} \label{eq11}	
&\,&\expect{ \cos^{k} \theta}(t,t_0,\boldsymbol{\epsilon},\boldsymbol{\delta})= \\
&\, & \!\sum_{\boldsymbol{n} \in \mathbb{Z}^{s}} \!C_{\boldsymbol{n}}(t,\boldsymbol{\epsilon}) \prod_{j=1}^{s} \epsilon_{j}^{|n_{j}|} \cos[\boldsymbol{n}\cdot\boldsymbol{\delta}\!+\!\omega t_0 \boldsymbol{n}\cdot\boldsymbol{q} \!+\!\Theta_{\boldsymbol{n}}(t,\boldsymbol{\epsilon})], \nonumber 
\end{eqnarray}
where $\boldsymbol{q}=\{q_1,\cdots,q_s\}$, $\boldsymbol{\epsilon}=\{\epsilon_1,\cdots,\epsilon_s\}$,  
${\boldsymbol{\delta}}=\{\delta_1,\cdots,\delta_s\}$, and $C_{\boldsymbol{n}}(t,\boldsymbol{\epsilon})$ and $\Theta_{\boldsymbol{n}}(t,\boldsymbol{\epsilon})$ 
are both even functions of each $\epsilon_j$.
The $t_0$-averaged expectation values 
satisfy~\cite{Casado} 
\begin{eqnarray} \label{eq17}
&&\expect{\expect{ \cos^{k} \theta}}(t,\boldsymbol{\epsilon},\boldsymbol{\delta})= \\
&&C_{\boldsymbol{0}}(t,\boldsymbol{\epsilon})+\!\sum_{\boldsymbol{n} \in \mathbb{S}} \!C_{\boldsymbol{n}}(t,\boldsymbol{\epsilon}) \prod_{j=1}^{s} \epsilon_{j}^{|n_{j}|} \cos[\boldsymbol{n}\cdot\boldsymbol{\delta} \!+\!\Theta_{\boldsymbol{n}}(t,\boldsymbol{\epsilon})], \nonumber
\end{eqnarray} 
where 
\begin{equation}
\mathbb{S}=\{\boldsymbol{n} \in \mathbb{Z}^{s}: \boldsymbol{n} \cdot \boldsymbol{q}=0\}
\label{eq:Diophantine}
\end{equation}
denotes the set of nonzero solutions of the Diophantine equation $\boldsymbol{n} \cdot \boldsymbol{q}=0$, whose leftmost nonzero component is positive \cite{Cuesta}. 

If only the interaction of the electric field with the  electric dipole moment  is included
in Hamiltonian~\eqref{eq:molecular_Hamiltonian},  
 $s=2$, and 
$\boldsymbol{q}=\{q_1,q_2\}$, $\boldsymbol{\epsilon}=\{\epsilon_1,\epsilon_2\}$, and 
${\boldsymbol{\delta}}=\{\delta_1,\delta_2\}$~\cite{quintero:2010,Cuesta}.
The  Diophantine equation is 
$n_1 q_1+ n_2 q_2=0$.
The $t_0$-averaged expectation value~\eqref{eq:t0_average} can be written as
\begin{eqnarray}
\label{eq:expect_fit_general} 
&&\expect{\expect{\cos^k\theta}}(t,\epsilon_1,\epsilon_2,\delta_1,\delta_2)= \\
&&\sum_{j=0}^{+\infty}|C_j(t,\epsilon_1,\epsilon_2)| \left(\epsilon_1^{q_2} \epsilon_2^{q_1}
\right)^j\cos\big[j\xi_{12}+\Theta_j(t,\epsilon_1,\epsilon_2)\big],\nonumber
\end{eqnarray}
where $\xi_{12}=(q_1\delta_2-q_2\delta_1)$ and 
$\Theta_0(t,\epsilon_1,\epsilon_2)=0$~\cite{Casado}.  
Due to the inversion of the electric field direction symmetry~\eqref{eq:sym_t0_force}, it holds that:
i) the series~\eqref{eq:expect_fit_general} includes  only even terms for  $k$ even  if
$q_1+q_2$ is odd, otherwise all the terms contribute;
ii)~\autoref{eq:expect_fit_general} includes only odd  ones for $k$ odd if $q_1+q_2$ is odd; 
and iii) if  $q_1+q_2$ is   even,
$\expect{\expect{\cos^k\theta}}=0$ with $k$ odd, and for $k=1$, the molecule is not  oriented.

By including also the interaction with the polarizability, 
six  harmonics appear in the Hamiltonian, \ie,  $s=6$  
in the set~\eqref{eq:Diophantine} of non-zero solutions of the Diophantine equation,
which  reads 
\begin{equation}
\label{eq:diophantine}
(n_1+ 2 n_3+n_5-n_6) q_1+ (n_2+2 n_4+n_5+n_6) q_2=0,
\end{equation}
and  $\mathbb{S}= \{n_1=- 2 n_3-n_5+n_6-m q_2, n_1>0, n_2=-2 n_4-n_5-n_6+m q_1, 
(m, n_3, n_4, n_5, n_6) \in \mathbb{Z}^{5} \}$. As a consequence, $\boldsymbol{n} \cdot \boldsymbol{\delta}=m \xi_{12}$. 
Using this result,~\autoref{eq17} can be rewritten as
\begin{eqnarray}\label{eq17A}
&&\expect{\expect{ \cos^{k} \theta}}(t,\epsilon_1,\epsilon_2,\delta_1,\delta_2)=C_{\boldsymbol{0}}(t,\epsilon_{1},\epsilon_{2})+\\
\nonumber && \!\sum_{(\boldsymbol{n},m) \in \mathbb{S}} \!C_{\boldsymbol{n}}(t,\epsilon_{1},\epsilon_{2}) \epsilon_{1}^{|x_{\boldsymbol{n}}|} 
\epsilon_{2}^{|y_{\boldsymbol{n}}|} 
  \cos[m \xi_{12} \!+\!\Theta_{\boldsymbol{n}}(t,\epsilon_{1},\epsilon_{2})],  
\end{eqnarray} 
where $x_{\boldsymbol{n}}$, $y_{\boldsymbol{n}}$ and $m$ are integers determined not only by the solutions of the Diophantine equation, but also by the symmetries.  Indeed, 
 the symmetry  ~\eqref{eq:sym_t0_force} implies that $|x_{\boldsymbol{n}}|+|y_{\boldsymbol{n}}|$ has the same parity as $k$. 
 Thus, this expectation value~\eqref{eq17A} is rewritten as 
 \begin{eqnarray}\label{eq4}
 &&\expect{\expect{ \cos^{k} \theta}}(t,\epsilon_1,\epsilon_2,\delta_1,\delta_2)= 
 \\
\nonumber
&&\sum_{j=0}^{+\infty} \!{ \cal C}_j(t,\epsilon_1,\epsilon_2) \cos\left[j\xi_{12} \!+\!\varphi_j(t,\epsilon_1,\epsilon_2)\right]  \text{,} \end{eqnarray} 
 with ${ \cal C}_j(t,-\epsilon_1,-\epsilon_2)=(-1)^k { \cal C}_j(t,\epsilon_1,\epsilon_2)$. 
 Furthermore, due to the symmetry on 
 the phases~\eqref{eq:sym_t0_d1_d2}
 two cases  can be  distinguished
according to the parity of $q_1+q_2$.
(i) If $q_1+q_2$ is an even integer, $q_1$ and $q_2$ are both odd integer numbers because  $\text{gcd}(q_1,q_2)=1$. 
As a consequence of  the symmetry~\eqref{eq:sym_t0_d1_d2}, the identity 
$\expect{\expect{ \cos^{k} \theta}}=(-1)^k\,\expect{\expect{ \cos^{k} \theta}}$ is obtained. 
Therefore,  the expectation values is zero for $k$ odd and satisfies~\eqref{eq4} for $k$ even.
(ii) If $q_1+q_2$ is an odd integer, $q_1$ and $q_2$ have different parity.
Due to  the  symmetry~\eqref{eq:sym_t0_d1_d2},  
for odd or even values of $k$ in~\eqref{eq4},  only 
odd or even terms contribute to the  corresponding expansion, respectively,  \ie,
\begin{eqnarray} \label{fit1}
&&\expect{\expect{ \cos^{2k+1} \theta}}(t,\epsilon_1,\epsilon_2,\delta_1,\delta_2)= \\ \nonumber &&\sum_{\stackbin[(j\ odd)]{}{{j=1}}}^{+\infty} \!{ \cal C}_j(t,\epsilon_1,\epsilon_2)\cos\left[j\xi_{12} \!+\!\varphi_j(t,\epsilon_1,\epsilon_2)\right]  \text{,}
\end{eqnarray}	
and 
	\begin{eqnarray} \label{fit2}
&&	\expect{\expect{ \cos^{2k} \theta}}(t,\epsilon_1,\epsilon_2,\delta_1,\delta_2)= \\ \nonumber
&&\sum_{\stackbin[(j\ even)]{}{{j=0}}}^{+\infty} \!{\cal C}_j(t,\epsilon_1,\epsilon_2)\cos\left[j\xi_{12} \!+\!\varphi_j(t,\epsilon_1,\epsilon_2)\right] \text{.}
	\end{eqnarray} 
	
 When the interaction between the electric field and hyperpolarizability is also included in the Hamiltonian,  
$s=14$ in the set of  non-zero solutions~\eqref{eq:Diophantine} of the Diophantine equation,
which reads
$[n_1+ 2 (n_3+n_{11}-n_{12})+n_5-n_6+n_7+ 3 n_9+n_{13}-n_{14}] q_1+  
[n_2+2(n_4+n_{13}+n_{14})+n_5+n_6+n_8+3 n_{10}+ n_{11}+n_{12}] q_2=0$.
Therefore, $\mathbb{S}= \{n_1=- 2(n_3+n_{11}-n_{12})-n_5+n_6-n_7-3 n_9-n_{13}+n_{14}-m q_2, n_1>0, n_2=-2 (n_4+n_{13}+n_{14})-n_5-n_6-n_8-3 n_{10}-n_{11}-n_{12}+m q_1, 
(m, n_3, n_4, \cdots,n_{14}) \in \mathbb{Z}^{13} \}$. As a consequence, $\boldsymbol{n} \cdot \boldsymbol{\delta}=m \xi_{12}$ is also satisfied. 
In this case,  expression~\eqref{eq17A} is also obtained, and  a similar symmetry analysis transforms it  into
 the formulas~\eqref{fit1} and~\eqref{fit2} if $q_1+q_2$ is an odd integer. 
Finally, note  that~\eqref{eq:expect_fit_general} can also be rewritten as~\eqref{fit1} and~\eqref{fit2} if $q_1+q_2$ is an odd integer.  

\bibliographystyle{apsrev4-1}  

\bibliography{references}

\end{document}